\documentclass[aps,prb,superscriptaddress,preprint]{revtex4-1}

\usepackage{todonotes}

\usepackage{times}
\usepackage{amsmath}
\usepackage{amsfonts}
\usepackage{amssymb}
\usepackage{epsfig}
\usepackage{siunitx}
\usepackage{gensymb}

\usepackage{float}
\usepackage{ifthen}
\usepackage{xspace}
\usepackage{relsize}

\usepackage{algorithm}
\usepackage{algorithmic}

\begin{document}

\title{Microwave photons emitted by fractionally charged quasiparticles}

\author{R. Bisognin$^{1}$, H. Bartolomei$^{1}$, M. Kumar$^{1}$,  I. Safi$^{2 }$, J.-M. Berroir$^{1 }$,  E. Bocquillon$^{1 }$, B.
Pla\c{c}ais$^{1 }$,   A. Cavanna$^{3}$, U.
Gennser$^{3 }$, Y. Jin$^{3 }$, and G. F\`{e}ve$^{1 \ast}$ \\
\normalsize{$^{1}$Laboratoire
de Physique de l\textquoteright Ecole normale sup\'erieure, ENS, Universit\'e
PSL, CNRS, Sorbonne Universit\'e, Universit\'e Paris-Diderot, Sorbonne Paris
Cit\'e, Paris, France}\\
\normalsize{$^{2}$ Laboratoire de Physique des Solides, Universit\'{e} Paris-Saclay, 91405 Orsay France}\\
\normalsize{$^{3}$ Centre de Nanosciences et de Nanotechnologies (C2N), CNRS, Univ. Paris Sud, Universit\'{e} Paris-Saclay,}\\ \normalsize{
91120 Palaiseau, France.}\\
\normalsize{$^\ast$ To whom correspondence should
be addressed; E-mail:  feve@lpa.ens.fr.}\\
}

\begin{abstract}
Strongly correlated low-dimensional systems can host exotic elementary excitations carrying a fractional charge $q$ and potentially obeying anyonic statistics. In the fractional quantum Hall effect, their fractional charge has been successfully determined owing to low frequency shot noise measurements. However, a universal method for sensing them unambiguously and unraveling their intricate dynamics was still lacking. Here, we demonstrate that this can be achieved by measuring the microwave photons emitted by such excitations when they are transferred through a potential barrier biased with a dc voltage $V_{\text{dc}}$. We observe that only photons at frequencies $f$ below $qV_{\text{dc}}/h$ are emitted. This threshold provides a direct and unambiguous determination of the charge $q$, and a signature of exclusion statistics. Derived initially within the Luttinger model, this feature is also predicted by universal non-equilibrium fluctuation relations which agree fully with our measurements. Our work paves the way for further exploration of anyonic statistics using microwave measurements.
\end{abstract}

\maketitle

As revealed by Shottky\cite{Schottky1918}, the random transmission of charged carriers in an electrical circuit leads to low frequency fluctuations of the current that are directly proportional to the transmission probability and the charge of the carriers. In the fractional quantum Hall effect\cite{Tsui1982}, low frequency noise measurements\cite{dePicciotto1997,Saminadayar1997,Reznikov1999} have been crucial to evidence fractionally charged quasiparticles\cite{Laughlin83}. However, they suffer from several drawbacks. Firstly, the noise depends not only on the nature of the charge carriers, but also on the scattering properties of the conductor. For example, the low frequency noise generated by electrons propagating through a coherent diffusive conductor \cite{Henny1999} is the same as the one generated by the transfer of quasiparticles of charge $1/3$ through a weak potential barrier\cite{dePicciotto1997,Saminadayar1997}, even though the origin of the noise is very different. Secondly, for specific filling factors of the fractional quantum Hall effect, these measurements do not provide an unambiguous determination of the charge, which is found to depend on external parameters such as electronic temperature or conductor transmission\cite{Chung2003,Bid2009,Dolev2010}. Finally, being intrinsically low frequency, they cannot probe the characteristic frequency scales of quasiparticle transfer in contrast with high frequency measurements\cite{Gabelli2009}.

In this paper, we report on the measurement of high frequency noise (at frequency $3.5 \leq f \leq 8$ GHz) generated by the transfer of fractionally charged quasiparticles through a quantum point contact (QPC) in the weak backscattering regime. At frequencies $hf \gg k_{\text{B}} T$ (where $T\approx 50$ mK is the electronic temperature), the current fluctuations result from the emission of microwave photons collected in a measurement transmission line weakly coupled to the sample \cite{schoelkopf97,Zakka-Bajjani2007}. A continuous stream of quasiparticles impinging on the QPC is generated by applying a dc voltage $V_{\text{dc}}$ to the sample. We observe that photon emission resulting from quasiparticle transfer only occurs if the photon energy $hf$ is smaller than the energy $qV_{\text{dc}}$ of the quasiparticles, thereby revealing their exclusion statistics\cite{Haldane1991}: quasiparticles cannot emit photons at energies higher than $qV_{\text{dc}}$ as no empty states are available at energy $qV_{\text{dc}}-hf$. At fixed measurement frequency $f$, it results in a threshold, $V_{\text{dc}} \geq V_0=hf/q$, for the observation of photon emission. By measuring the emission threshold as a function of frequency, we provide a direct measurement of the fractional charge without any need for noise calibration or knowledge of the conductor scattering properties. This emission threshold, derived initially within the Luttinger model\cite{Chamon95,Bena2007}, is also predicted by universal non-equilibrium fluctuation relations\cite{Safi2011,Roussel2016} which agree perfectly with our measurement and do not rely on any assumption on the microscopic description of the system. This agreement establishes the role of the fractional charge in the dynamics of quasiparticle transfer through the characteristic frequency scale $f_{\text{J}}=qV_{\text{dc}}/h$ called the Josephson frequency\cite{Chamon95,Bena2007,Kapfer2018} in analogy with the Josephson relation $f_{\text{J}}=2e V_{\text{dc}}/h$ in superconductors. This work opens the way to the exploration of anyonic statistics\cite{Wilczek1982} from frequency dependent noise measurements\cite{Kim2005,Bena2006,Carrega2012}. They are complementary to recent experiments\cite{Kapfer2018} of photo-assisted low frequency noise in the fractional quantum Hall regime\cite{Crepieux2004} with the advantage of not requiring to shine microwaves on the sample.

\section*{Results}

\begin{figure}[hhh!]
\includegraphics[width=0.8
\columnwidth,keepaspectratio]{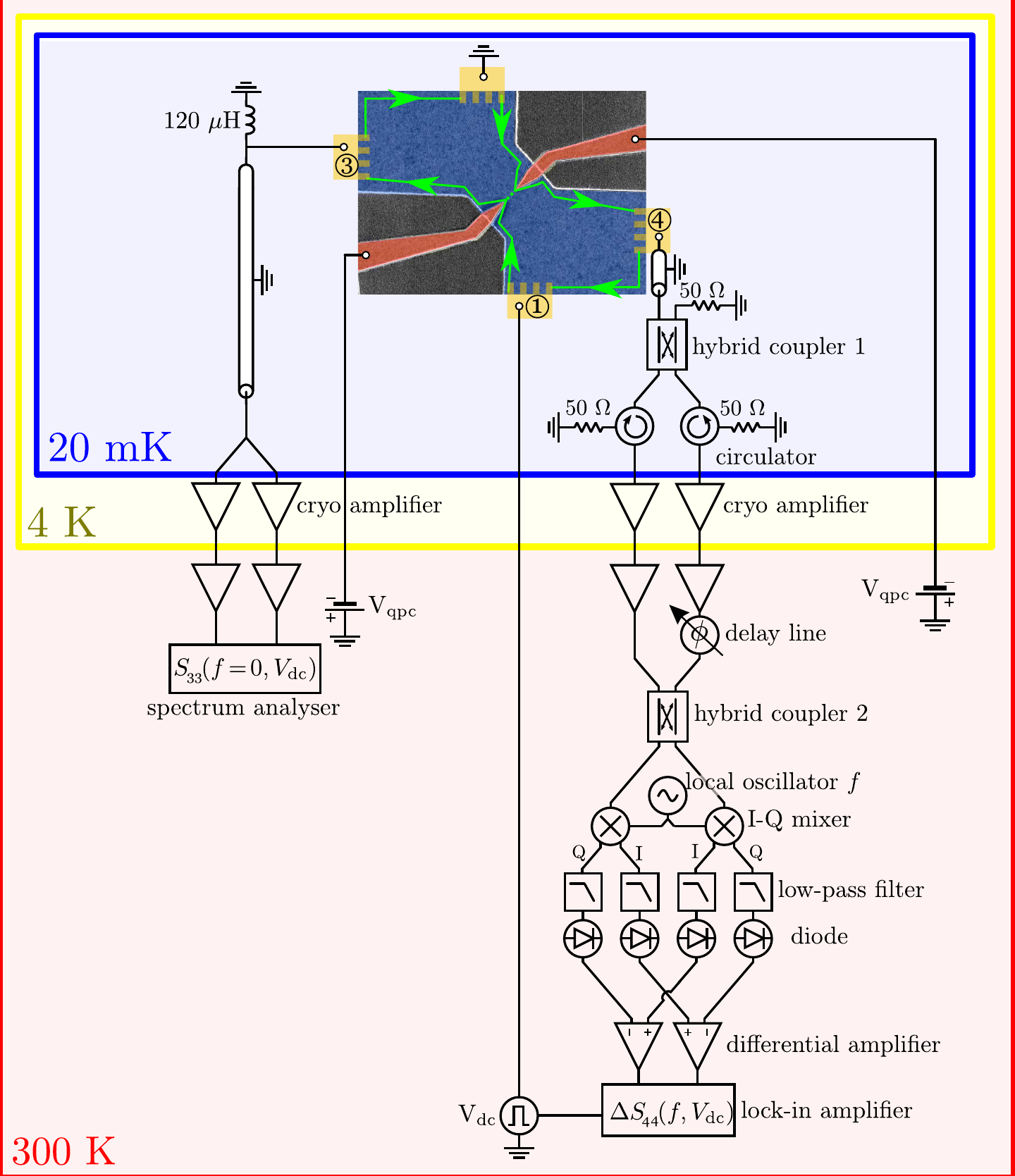}
\caption{ Experimental setup. The low frequency shot noise is measured at output 3 of the sample and high frequency one at output 4.  A square pulse drive $V_{\text{dc}}(t)$ modulates the applied voltage on input between $0$ and the variable amplitude $V_{\text{dc}}$ at frequency $234$ Hz for lock-in detection of the high frequency noise (see Methods). The microwave noise emitted by the sample is split, amplified and recombined\cite{Parmentier2011} at the output $0\degree$ of hybrid coupler 2. The noise power on the I and Q quadratures at the measurement frequency $f$ are measured using I/Q mixers followed by diodes in a $1.5$ GHz bandwidth defined by low pass filters. Low frequency differential amplifiers measure the power difference between the two outputs of hybrid coupler 2 in order to  subtract the amplifier noise (see Methods). The emitted noise on the I and Q quadratures are then measured using lock-in detection at the modulation frequency $234$ Hz and averaged together as the emitted noise is the same for the two quadratures. \label{fig1} }
\end{figure}

{\bf Sample and experimental setup.}
The experiment is performed on a high mobility two-dimensional electron gas in a GaAs/AlGaAs heterojunction (see Fig.\ref{fig1}). To compare our results between integer and fractional charges, we implement the experiment in the integer (filling factor $\nu=3$) and fractional ($\nu=4/3$ and $\nu=2/3$) quantum Hall regime. The dc current is generated by applying a dc bias to ohmic contact 1. Ohmic contact 3 is used to measure the differential conductance through the QPC and the low frequency excess current fluctuations (see Methods) generated by the scattering of quasiparticles at the QPC: $\Delta S_{33}(f=0,V_{\text{dc}})=S_{33}(f=0,V_{\text{dc}})-S_{33}(f=0,V_{\text{dc}}=0)$. Contact 4 is connected to a transmission line for high frequency excess noise measurements $\Delta S_{44}(f,V_{\text{dc}})= S_{44}(f,V_{\text{dc}})-S_{44}(f,V_{\text{dc}}=0)$ (see Methods).\\

\begin{figure}[hhh!]
\includegraphics[width=1
\columnwidth,keepaspectratio]{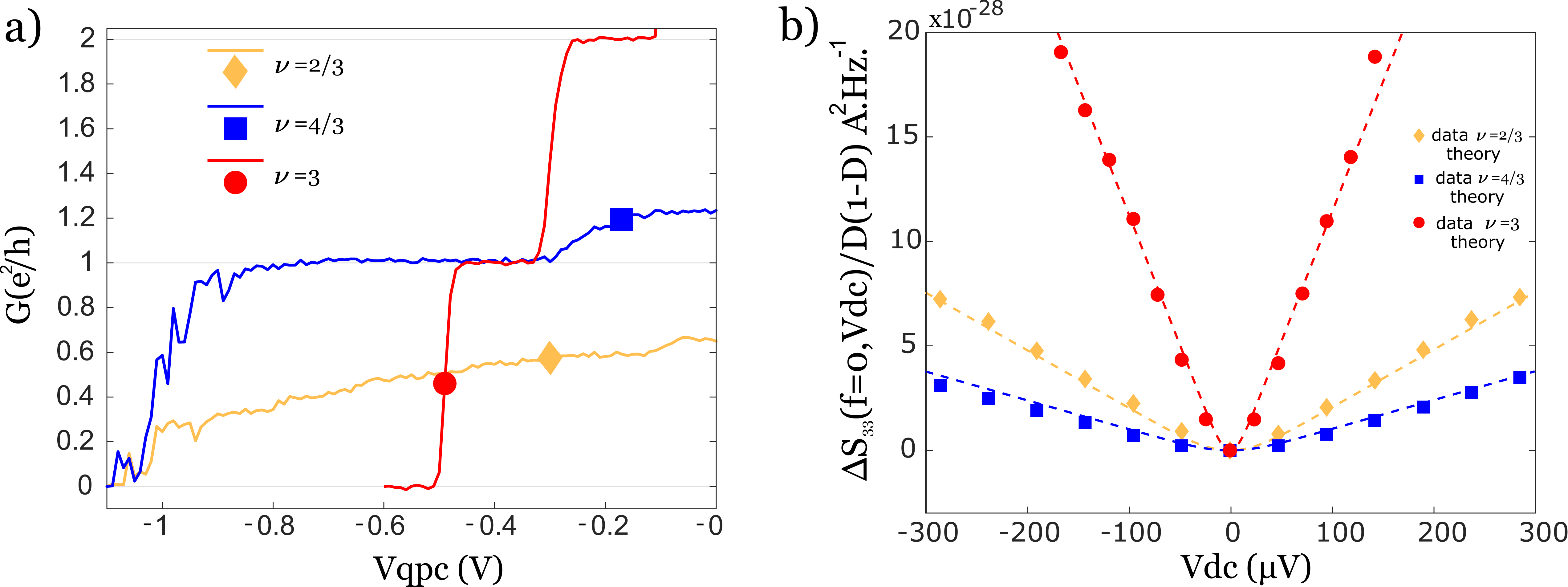}
\caption{Low frequency measurements. \textbf{a} Conductance through the QPC as a function of the QPC gate voltage for the three filling factors $\nu=3$ (red), $\nu=4/3$ (blue) and $\nu=2/3$ (yellow). The values of the QPC gate voltage used for noise measurements are shown by a colored disk ($\nu=3$), square ($\nu=4/3$) and diamond ($\nu=2/3$). \textbf{b} Measurements of $\Delta S_{33}(f=0,V_{\text{dc}})$ normalized by the factor $D(1-D)$. Dashed lines represent plots of Eq.(\ref{lfnoise}) using $q=e$, $g=\frac{e^2}{h}$ at $\nu=3$, $q=e/3$, $g=\frac{e^2}{3h}$ at $\nu=4/3$ and $q=e/3$, $g=\frac{2 e^2}{3 h}$ at $\nu=2/3$. \label{fig2} }
\end{figure}

{\bf Low frequency characterization.} Figure \ref{fig2}a presents the characterization of the conductance $G$ through the QPC for the three filling factors. At $\nu=3$, the current is carried by three edge channels of conductance $e^2/h$, we set the QPC at transmission $D=0.5$ of the outer channel ($G=0.5 \frac{e^2}{h}$) for the noise measurements. For $\nu=4/3$ we observe the transmission of two channels of conductance $e^2/h$ and $e^2/3h$. We set the QPC at $D=0.63$ for the $\nu=1/3$ channel ($G=\frac{e^2}{h} +0.63 \frac{e^2}{3h}$). For $\nu=2/3$ we set the QPC at $D=0.87$ ($G=0.87 \frac{2e^2}{3h}$). In all cases, $D$ varies weakly with $V_{dc}$ allowing us to assume a bias independent transmission in the analysis. Our measurements of $\Delta S_{33}(f=0,V_{dc})$ normalized by the Fano factor $D(1-D)$ are plotted in Figure \ref{fig2}b. The dashed lines represent predictions using the non-interacting formula\cite{Reznikov1999,Blanter2000,Martin2005,Feldman2017}:
\begin{equation}
    \frac{\Delta S_{33}(f=0,V_{\text{dc}})}{D(1-D)}=2qg_0 V_{\text{dc}}\left(\coth{\left(\frac{qV_{\text{dc}}}{2k_{\text{B}} T}\right)} - \frac{2k_{\text{B}} T}{qV_{\text{dc}}}\right),
    \label{lfnoise}\end{equation}
where $g_0$ is the conductance of the scattering channel ($g_0=\frac{e^2}{h}$ for $\nu=3$, $g_0=\frac{1}{3} \frac{e^2}{h}$ for $\nu=4/3$, and $g_0=\frac{2}{3} \frac{e^2}{h}$ for $\nu=2/3$).  Our measurements are consistent with the integer charge $q=e$ at $\nu=3$ (red dashed line) and the fractional charge $q=e/3$ at $\nu=2/3$ and $\nu=4/3$ (yellow and blue dashed line).\\

\begin{figure}[hhh!]
\includegraphics[width=0.45
\columnwidth,keepaspectratio]{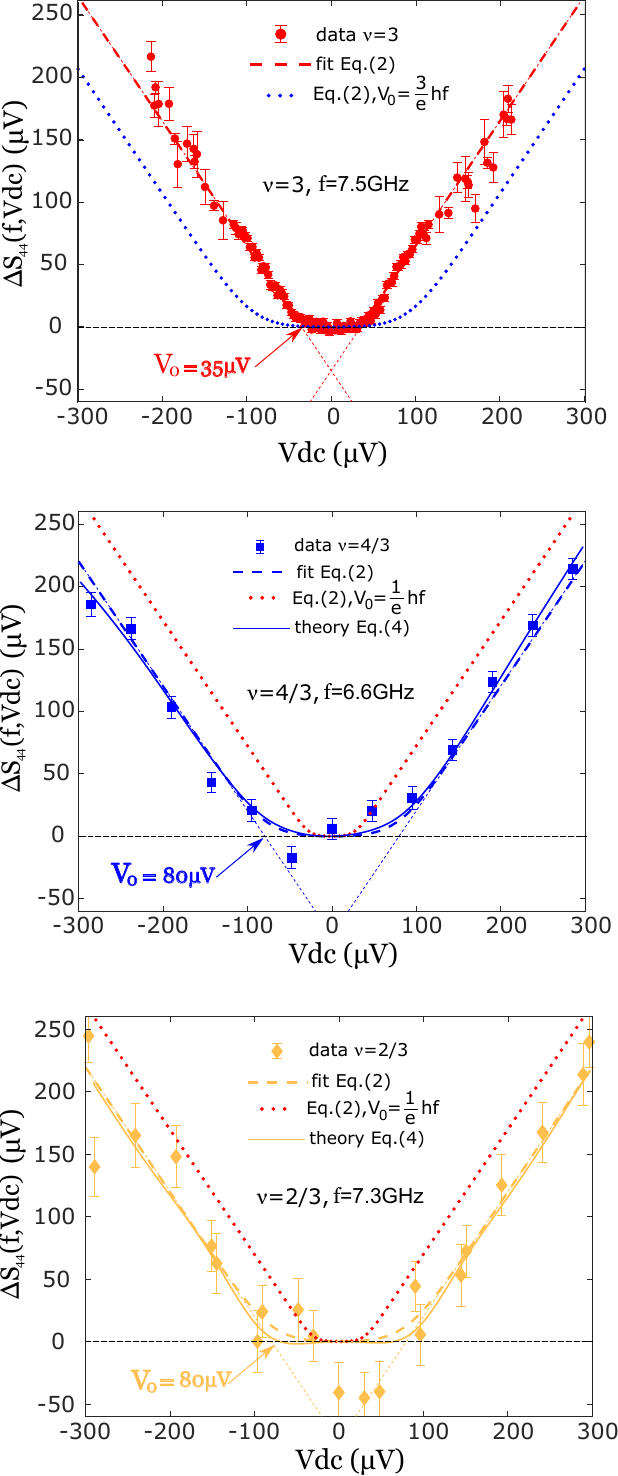}
\caption{High frequency noise measurements. Measurements of $\Delta S_{44}(f,V_{\text{dc}})$ for the three filling factors at $f \approx 7$ GHz. Fits by Eq.(\ref{HFclass}) with $V_{0}$ as the fitting parameter are plotted in dashed lines. $\Delta S_{44}$ is rescaled in Volts (see manuscript text). Comparisons with non-equilibrium FDR, Eq.(\ref{FDR}), using the measurements of $\Delta S_{44}(f=0,V_{\text{dc}})$ are represented in solid lines. Finally, the dotted lines represent plots of Eq.(\ref{HFclass}) with charge $q=e/3$ at $\nu=3$ (blue dotted line) and $q=e$ at $\nu=4/3$ and $\nu=2/3$ (red dotted line). Error bars are defined as standard error of the mean. \label{fig3} }
\end{figure}

\begin{figure}[hhh!]
\includegraphics[width=0.55
\columnwidth,keepaspectratio]{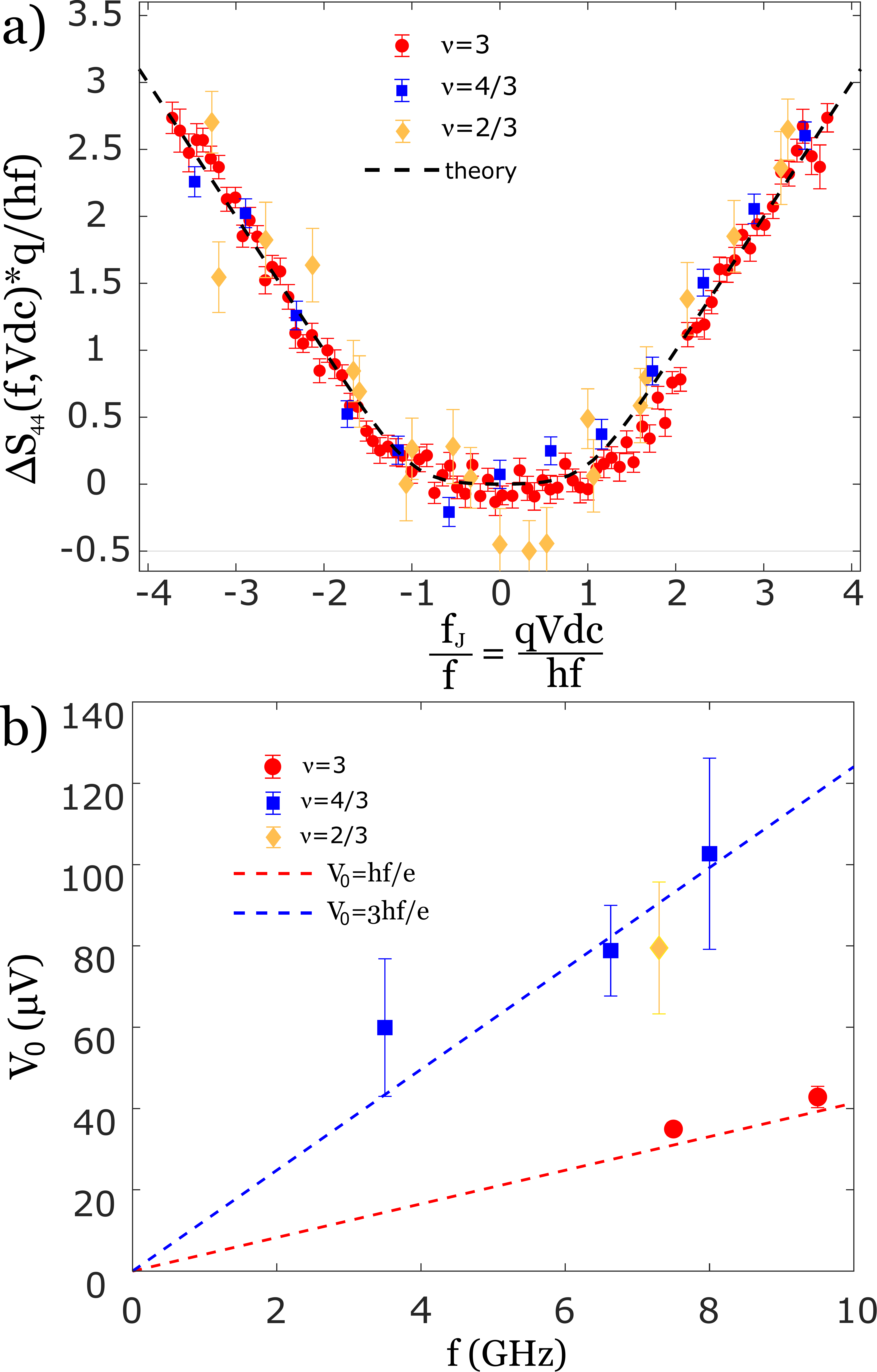}
\caption{The characteristic frequency for fractional quasiparticle transfer. \textbf{a} High frequency noise measurements $\frac{q\Delta S_{44}(f,qV_{\text{dc}}/(hf))}{hf}$  as a function of the ratio of the characteristic Josephson frequency $f_\text{J}=qV_{\text{dc}}/h$ with the measurement frequency $f$. We take $q=e$ for $\nu=3$ and $q=e/3$ for $\nu=4/3$ and $\nu=2/3$. The black dashed line is $\frac{x+1}{2}\coth{\frac{x+1}{2x_{T}}}+\frac{x-1}{2}\coth{\frac{x-1}{2x_{T}}}- \coth{\frac{1}{2x_{T}}}$ where $x_{T}=\frac{k_{\text{B}}T}{hf} \approx 0.15$ is the electronic temperature in renormalized units. Error bars are defined as standard error of the mean. \textbf{b} Emission threshold $V_{0}$ for the three filling factors as a function of the measurement frequency $f$. The blue dashed line is $V_0=3 hf/e$. The red dashed line is $V_0=hf/e$. Error bars represent the 68 percent confidence interval. \label{fig4} }
\end{figure}

{\bf High frequency noise measurements.} Having thus independently characterized the charge of the elementary excitations transferred at the QPC, we now turn to the high frequency noise measurements $\Delta S_{44}(f,V_{\text{dc}})$ plotted in Figure \ref{fig3} for $f\approx 7$ GHz. At high $V_{dc}$, we observe a linear dependence for all filling factors: $\Delta S_{44}(f,V_{\text{dc}}) = \alpha |V_{\text{dc}} - V_0|$. In order to compare measurements at different filling factors without the need for noise calibration, we scale the noise values on all plots by imposing a slope $\alpha =1$ at large bias. At small $V_{\text{dc}}$, the excess noise is approximately zero up to the voltage threshold $V_{0}$ where it starts to increase with a linear dependence. In order to extract a quantitative determination of $V_0$, we fit the experimental data with the non-interacting model\cite{Blanter2000,Martin2005} for the high frequency noise, using $V_0$ as the fitting parameter and imposing the same slope $\alpha =1$:
\begin{widetext}
\begin{equation}
\Delta S_{44}(f,V_{\text{dc}})  =  \frac{V_{\text{dc}}+V_{0}}{2}\coth{\left(\frac{q(V_{\text{dc}}+V_{0})}{2k_{\text{B}} T}\right)} + \frac{V_{\text{dc}}-V_{0}}{2}\coth{\left(\frac{q(V_{\text{dc}}-V_{0})}{2k_{\text{B}} T}\right)} -V_{0}\coth{\left(\frac{qV_{0}}{2k_{\text{B}} T}\right)}
\label{HFclass}
\end{equation}
\end{widetext}
The fits (dashed lines on Fig.\ref{fig3}) agree very well with the data, surprisingly even in the fractional case although Eq.(\ref{HFclass}) has a priori no reason to be valid in this situation. We find $V_0=35 \pm 1.6\mu V$ at $\nu=3$,  $V_0=80 \pm 11\mu V$ at $\nu=4/3$ and $V_0=80 \pm 17\mu V$ at $\nu=2/3$. This difference is related to the difference in the quasiparticle charge $q$. The dynamics of quasiparticle transfer are governed by the frequency scale $f_\text{J}=q V_{\text{dc}}/h$, and photon emission at frequency $f$ only occurs for $f_\text{J} \geq f$. From our determination of $V_0$, we obtain $q=0.89 \pm 0.04 e$ at $\nu=3$, $q=0.34 \pm 0.05 e$ at $\nu=4/3$ and $q=0.38 \pm 0.08 e$ at $\nu=2/3$ in reasonable accordance with the low-frequency determination of $q$. The characteristic frequency scale $f_\text{J}=q V_{\text{dc}}/h$ can be highlighted by rescaling the data by the factor $q/(hf)$ on the vertical and horizontal axes, taking $q=e$ for $\nu=3$ and $q=e/3$ for $\nu=4/3$ and $\nu=2/3$. As can be seen from Figure \ref{fig4}a, all data collapse, within experimental error bars, on the same trace (black dashed line) given by Eq.(\ref{HFclass}) with a threshold $f_\text{J}/f=1$. In order to check the consistency of our results for various measurement frequencies $f$, we plot in Figure \ref{fig4}b the emission thresholds extracted from all our data points as a function of $f$ (see Supplementary Note 3 for additional measurements). In the integer case, the measured thresholds $V_0$ fall on the red dashed line representing electron transfer: $V_0=hf/e$. In contrast, the thresholds measured in the fractional case fall on the blue dashed line $V_0=3 hf/e$ corresponding to quasiparticle transfer with charge $q=e/3$.\\

{\bf Comparison with non-equilibrium fluctuation relations.} Finally, by measuring simultaneously the low and high  frequency noises,  we can compare our measurements in the strongly interacting case with parameter free theoretical predictions. Using non-equilibrium fluctuation dissipation relations\cite{Roussel2016} (FDR) in the weak backscattering limit (see Methods), the high frequency backscattering noise $\Delta S_{\text{b}}(f,V_{\text{dc}})$ can be related to its low frequency value $\Delta S_{b}(f=0,V_{\text{dc}})$ through the characteristic frequency scale $f_\text{J}=qV_{\text{dc}}/h$:

\begin{equation}
\Delta S_{\text{b}}(f,V_{\text{dc}})  = \frac{\Delta S_{\text{b}}(f=0,V_{\text{dc}}+hf/q) + \Delta S_{\text{b}}(f=0,V_{\text{dc}}-hf/q)}{2}
\label{FDR}
\end{equation}

Eq.(\ref{FDR}) does not rely on any assumption on the nature of the weak backscattering (local or extended), on the nature of interactions, or on the microscopic description of edge channels. The only assumption is that a single backscattering process with a given charge $q$ needs to be considered, such that a single characteristic frequency scale $f_\text{J}=qV_{\text{dc}}/h$ governs the  dynamics of charge transfer. The experimentally accessible quantity $\Delta S_{44}$ differs from $\Delta S_b$ by a correction related to the non-linearity of the backscattering current (see Methods). However,  the non-linearity we measure is small, and we can assume (up to a correction $<5\%$, see Methods) that $\Delta S_{44}$ also satisfies Eq.(\ref{FDR}):

\begin{equation}
\Delta S_{44}(f,V_{\text{dc}}) =\frac{\Delta S_{44}(f=0,V_{\text{dc}}+hf/q) + \Delta S_{44}(f=0,V_{\text{dc}}-hf/q)}{2}
\label{deltS44}
\end{equation}

The blue and yellow solid lines in Fig.\ref{fig3} represent the predictions of Eq.(\ref{deltS44}) using our measurements of $\Delta S_{33}(f=0,V_{\text{dc}})=\Delta S_{44}(f=0,V_{\text{dc}})$. The agreement between the data and the non-equilibrium FDR is excellent, providing a stringent verification of its validity in this strongly interacting case, and justifying a-posteriori the agreement with the non-interacting theory of Eq.(\ref{HFclass}) up to a renormalization of the frequency scale $qV_{\text{dc}}/h$ by the fractional charge $q=e/3$.

\section*{Discussion}
To conclude, we have observed that fractionally charged quasiparticles scattered at a QPC could only generate photons at frequency $f \leq f_\text{J}=qV_{\text{dc}}/h$, providing a signature of their exclusion statistics. Using non-equilibrium FDR, we have explicitly connected high and low frequency measurements through the single characteristic frequency scale $f_\text{J}$, allowing for a determination of the fractional charge free from calibration or experimental parameters. A direct extension of this work is the measurement of high frequency noise for $\nu=2/5$ in the Jain sequence where charge transfer should result from different processes associated to different charges\cite{Ferraro2008} which should be revealed by different characteristic frequencies\cite{Ferraro2014}. In the $\nu=5/2$ case, high frequency noise could be used to probe the nature of the ground state and to distinguish between different descriptions\cite{Bena2006,Carrega2012}. Finally, more complex geometries involving additional QPC\cite{Safi2001} could also be used to investigate anyon statistics. The measurements presented in our work could be, for example, adapted to a T-junction geometry involving two QPC's backscattering fractional quasiparticles towards two distinct output edge channels\cite{Safi2001,Kim2005}. The current cross-correlations between these two outputs are the sum of two contributions, a direct term and an exchange term which directly probes the angle $\theta$ associated (in the abelian case) to the exchange of two anyons. The high frequency noise measurements we demonstrate in this work could then be used to distinguish these two contributions from their frequency dependence. In particular, the contribution of the exchange term is predicted\cite{Kim2005} to be dominant near the Josephson frequency such that bunching associated with a bosonic behavior ($\theta \leq \pi/2$) would be directly revealed by positive cross-correlations whereas antibunching associated with a fermionic behavior ($\theta \geq \pi/2$) would show up as negative cross-correlations at the measurement frequency $f=f_\text{J}$.

\section*{Methods}

\subsection{Sample and low-frequency characterization}
The sample is a GaAs/AlGaAs two dimensional electron gas of charge
density $n_{\text{s}}=\SI{1.9e15}{\per\square\meter}$ and mobility
$\mu=\SI{2.4e6}{\per\square\centi\meter\per\volt\per\second}$.
Low frequency noise measurements at output 3 of the sample are performed by measuring the cross-correlations between the voltages at the output of two cryogenic low-frequency amplifiers followed by room temperature amplifiers. The measurement frequency is set by the resonance frequency $f_{0}=1.1$ MHz of the $L C$ tank circuit connected to output 3 (see Fig.\ref{fig1}). The cross-correlations are done by a vector signal analyzer in a $200$ KHz bandwidth centered on $f_0$. The calibration of the low-frequency noise is performed by measuring the thermal noise of the resistance $R_{\nu}=h/(\nu e^2)$ seen from the tank circuit a as function of temperature. Measurements are performed at filling factors $\nu=3$ ($B=2.6$ T), $\nu=4/3$ ($B=5.9$ T) and $\nu=2/3$ ($B=12$ T). At $\nu=3$, the noise is generated by the partitioning of a single edge channel (outer channel) of quantized conductance $e^2/h$ set at transmission $0.5$ ($G=0.5 \frac{e^2}{h}$). The suppression of the noise at the conductance plateau $G=\frac{e^2}{h}$ supports the description of transport through three distinct channels. At $\nu=4/3$, we observe a conductance plateau at $G=1 \frac{e^2}{h}$ associated with a suppression of the noise (see Supplementary Note 1). This is consistent with the successive transmission of two channels of respective conductance of $\frac{e^2}{h}$ and $\frac{1}{3}\frac{e^2}{h}$ (although we do not reach the perfect transmission of the $\nu=1/3$ channel). When partitioning the $\nu=1/3$ channel, we observe the transfer of the fractional charge $q=e/3$. At $\nu=2/3$, we do not observe the suppression of the noise on the conductance plateau $G=\frac{1}{3}{e^2}{h}$ leading us to consider the following description of a single channel of transmission varying from $0$ to $1$ when the conductance varies from $0$ to $\frac{2}{3}{e^2}{h}$. For transmissions $D\geq 0.6$, the transferred charge is given by $q=e/3$ (see Supplementary Note 1).

\subsection{High-frequency noise measurements}
 For high frequency noise measurements, a broadband coaxial cable connects output 4 to the hybrid coupler 1 which splits the signal towards two output ports\cite{Parmentier2011}. Each output is then amplified by low noise rf cryogenic amplifiers followed by room temperature amplifiers. A circulator is placed between each cryogenic amplifier and the sample to avoid microwave irradiation from the amplifiers. In order to subtract the amplifier noise, the two lines are recombined at room temperature using a second hybrid coupler. The phase difference between the two arms is tuned using phase shifters, such that the rf signal emitted by the sample is fully transmitted at the output $0\degree$ of hybrid coupler number 2. The amplifier noise is on the contrary equally divided between both outputs, and can be subtracted by measuring the power difference between the two outputs of the hybrid coupler. In order to vary the measurement frequency $f$ the signal is down-converted using two I/Q mixers mixing the signal generated by a microwave source at frequency $f$ and the signal coming out of the room temperature hybrid coupler. The power on each quadrature is then measured using diodes in a bandwidth of $1.5$ GHz set by low-pass filters. The output of the diodes are then connected to differential amplifiers in order to subtract the contribution of the amplifiers. Although this technique allows us to suppress most of the amplifier noise, the remaining noise floor is still larger than the signal $\Delta S_{44}(f,V_{\text{dc}})$. We thus add a low-frequency modulation to $\Delta S_{44}(f,V_{\text{dc}})$ by applying a square voltage $V_{\text{dc}}(t)$ at input $1$ of the sample between $0$ V and $V_{\text{dc}}$ at a modulation frequency of $234$ Hz. The differential amplifiers are then connected to the two inputs of a lock-in amplifier for the measurement of the two quadratures of the emitted noise: $\Delta S_{44,I}(f,V_{\text{dc}})$ and $\Delta S_{44,Q}(f,V_{\text{dc}})$. As the emitted noise is isotropic, and does not depend on the measured I/Q quadrature, our measurements of $\Delta S_{44}(f,V_{\text{dc}})$ are obtained by averaging together $\Delta S_{44,I}(f,V_{\text{dc}})$ and $\Delta S_{44,Q}(f,V_{\text{dc}})$. The noise sensitivity reached with this setup can be compared to the equivalent temperature variation of a $50$ $\Omega$ resistor. At $\nu=4/3$ our resolution (size of error bars) is a noise temperature of a few $\mu$K (the typical noise temperature at maximum voltage $V_{\text{dc}}\approx 300$ $\mu$V is approximately $50$ $\mu$K). At high magnetic field (filling factor $\nu=1$ and above), measurements using a vector network analyzer show that high frequency microwaves are attenuated when propagating though the sample. As a consequence, the microwave photons generated at the quantum point contact are also attenuated before reaching the measurement transmission line. This results in a smaller high frequency noise signal at $\nu=2/3$ compared to $\nu=4/3$, which explains the larger size of the error bars at $\nu=2/3$.

\subsection{Non-equilibrium fluctuation dissipation relations}

To establish the non-equilibrium FDR used in this work, we consider the following time-dependent Hamiltonian:
\begin{equation}
	\label{Hamiltonian}
	\mathcal{H}(t)\!\! = \!\! \mathcal{H}_0 +\!\! \;e^{-i 2 \pi f_\text{J} t} \hat{A}+ e^{i 2 \pi f_\text{J} t} \;\hat{A}^{\dagger} ,
\end{equation}
where $\mathcal{H}_0$ can describe edges at arbitrary filling factors, with mutual Coulomb interactions of arbitrary range and inhomogeneous form, and where the operator $\hat{A}$ describes the scattering process between the two edges, which can be non-local and spatially extended. In particular, our approach does not require to assume a description in term of chiral Luttinger liquid for charge transfer across the QPC. We only assume a dominant scattering process of a given charge $q$, such that the Josephson-type frequency $f_\text{J}$ accounts for the effect of the dc voltage $V_{\text{dc}}$ on the fractionally charged quasiparticles through: $f_\text{J}=qV_{\text{dc}}/h$. The backscattering current operator $\hat{I}_{\text{b}}$ is then given by $\hat{I}_{\text{b}}(t) = \frac{-i q}{\hbar}\left[e^{-i 2 \pi f_\text{J} t}\;\hat{A}-e^{i 2 \pi f_\text{J} t}\; \hat{A}^{\dagger}\right]$.

The high frequency noise of the backscattering current $S_{\text{b}}(f,V_{\text{dc}})$ can be decomposed into its symmetric $S_{\text{b}}^{\text{sym}}(f,V_{\text{dc}})$ and antisymmetric $S_{\text{b}}^{\text{asym}}(f,V_{\text{dc}})$ parts with respect to the measurement frequency $f$. These two quantities obey the following non-equilibrium FDR\cite{Safi2011,Roussel2016}:
 \begin{eqnarray}
 S_{\text{b}}^{\text{asym}}(f,V_{\text{dc}}) & = & - h f \;\text{Re} \left( G_{\text{b}}(f,V_{\text{dc}}) \right) \\
 2 S_{\text{b}}^{\text{sym}}(f,V_{\text{dc}}) & = & q \coth{\left(\frac{qV_{\text{dc}}+hf}{2k_{\text{B}} T}\right)} I_{\text{b}}(V_{\text{dc}}+hf/q) +q \coth{\left(\frac{qV_{\text{dc}}-hf}{2k_{\text{B}} T}\right)} I_{b}(V_{\text{dc}}-hf/q) \nonumber\\ \label{Stt}
\end{eqnarray}
where $G_{\text{b}}(f,V_{\text{dc}})$ is the finite frequency admittance through the QPC and where Eq.(\ref{Stt}) is only valid in the weak backscattering regime. Using Eq.(\ref{Stt}) for $f=0$ and $f$ finite, we obtain Eq.(\ref{FDR}) of the main text which connects $\Delta S_{\text{b}}^{\text{sym}}(f,V_{\text{dc}}) $ to $\Delta S_{\text{b}}^{\text{sym}}(f=0,V_{\text{dc}}) $ through the characteristic frequency scale $f_\text{J}$.

In the experiment, we do not directly measure the noise of the backscattered current but rather current noises at the output ohmic contacts 3 and 4. Using again non-equilibrium FDR, we first notice that by measuring the excess noise, the asymmetric part of the measured noise vanishes. Indeed,
$S_{44}^{\text{asym}}(f,V_{\text{dc}})= -h f \text{Re}[G_{44}(f,V_{\text{dc}})]$\cite{Safi2011}, where $G_{44}(f,V_{\text{dc}})$ is the differential conductance at frequency $f$ and voltage $V_{\text{dc}}$ seen from the contact $4$. Here, thanks to chirality and the quantized conductance of edge channels, we have $G_{44}(f,V_{\text{dc}})= \nu e^{^2}/h$ independent of $V_{\text{dc}}$. Therefore $S_{44}^{\text{asym}}(f,V_{\text{dc}})$ does not depend on $V_{\text{dc}}$ and disappears from excess noise measurements. $ S_{44}^{\text{sym}}(f,V_{\text{dc}})$ can then be related to $S_{\text{b}}^{\text{sym}}(f,V_{\text{dc}})$ and  $G_{\text{b}}(f,V_{\text{dc}})$:
 \begin{eqnarray}
 S_{44}^{\text{sym}}(f,V_{\text{dc}}) & = &  S_{\text{b}}^{\text{sym}}(f,V_{\text{dc}}) -h f \coth{(\frac{hf}{2 k_{\text{B}} T})} \text{Re}(G_{\text{b}}(f,V_{\text{dc}})) \\
 \Delta S_{44}(f,V_{\text{dc}}) & = &  \Delta S_{\text{b}}^{\text{sym}}(f,V_{\text{dc}}) -h f \coth{(\frac{hf}{2 k_{\text{B}} T})} \text{Re}(\Delta G_{\text{b}}(f,V_{\text{dc}}))
 \end{eqnarray}
where $\Delta G_{\text{b}}(f,V_{\text{dc}})=G_{\text{b}}(f,V_{\text{dc}})-G_{\text{b}}(f,0)$ is the difference in the finite frequency admittance between zero bias and bias $V_{\text{dc}}$. For linear backscattering current (admittance independent of $V_{\text{dc}}$), we have $\Delta S_{44}(f,V_{\text{dc}}) = \Delta S_{\text{b}}^{sym}(f,V_{\text{dc}})$. In our situation, the backscattering current is not perfectly linear and the measured high and low frequency noises can be related up to a correction $\delta$ which can be evaluated:
\begin{eqnarray}
\Delta S_{44}(f,V_{\text{dc}}) & = &  \frac{\Delta S_{44}^{\text{sym}}(0,V_{\text{dc}}+hf/q)+ \Delta S_{44}^{\text{sym}}(0,V_{\text{dc}}-hf/q)}{2}-\delta \label{deltS44sup}\\
 \delta & = & c(f) -c(f=0) \\
  c(f) & = & h f \coth{(\frac{hf}{2 k_{\text{B}} T})} \text{Re}\left(G_{\text{b}}(f,V_{\text{dc}})-G_{\text{b}}(f,0)\right)
\end{eqnarray}
We do not directly measure the finite frequency admittance but, $\text{Re}\left[G_{\text{b}}(f,V_{\text{dc}})\right]$ can be expressed as a function of the weak backscattering current: $2 hf/q \text{Re}[G_{\text{b}}(f,V_{\text{dc}})]=I_{\text{b}}(V_{\text{dc}}+hf/q)-I_{\text{b}}(V_{\text{dc}}-hf/q)$. This allows for an evaluation of the correction $\delta$ from the measurement of $I_{\text{b}}(V_{\text{dc}})$. The evaluation of this correction is presented in the Supplementary Note 2. In the full range of $V_{\text{dc}}$ we estimate $\delta/\Delta S_{44}(f,V_{\text{dc}}) \leq 0.05$, such that we assume in the analysis $2 \Delta S_{44}(f,V_{\text{dc}})  =   \Delta S_{44}^{\text{sym}}(0,V_{\text{dc}}+hf/q)+ \Delta S_{44}^{\text{sym}}(0,V_{\text{dc}}-hf/q)$. Finally, due to symmetry between output contacts $3$ and $4$ we can use our low frequency noise measurements performed on output $3$ in Eq.(\ref{deltS44sup}) using $\Delta S_{44}(f=0,V_{\text{dc}})=\Delta S_{33}(f=0,V_{\text{dc}})$.

%

 \section*{Acknowledgements}
This work has been supported by ANR grants
``1shot reloaded'' (ANR-14-CE32-0017), and the ERC consolidator grant ``EQuO'' (No. 648236).

\section*{Author Contributions}

YJ fabricated the sample on GaAs/AlGaAs heterostructures grown by AC and UG. RB, HB and MK conducted the measurements. Theoretical models were developed by IS and simulations by RB, HB and GF. RB, HB, MK, JMB, EB, BP and GF participated to the analysis and the writing of the manuscript with inputs from IS, YJ, AC and UG. GF supervised the project.

\section*{Competing Interests}
The authors declare no competing interests.

\newpage

\newpage

 \newpage
 
\newpage

\end{document}